\documentclass[11pt,twoside]{article}

\usepackage{asp2006}
\usepackage{epsf}
\usepackage{psfig}
\usepackage{lscape}

\begin{document}
\title{Spectro-interferometric observations of interacting massive stars with VEGA/CHARA}   
\author{D. Bonneau, O. Chesneau, D. Mourard and P. Stee}   
\affil{UMR 6525 H. Fizeau, Univ. Nice Sophia Antipolis, CNRS, Observatoire de la C\^{o}te d'Azur, Av. N. Copernic, F-06130 Grasse, France}    

\begin{abstract} 
We obtained spectro-interferometric observations in the visible of $\beta$~Lyrae and $\upsilon$~Sgr using the instrument VEGA of the CHARA interferometric array. 
For $\beta$~Lyrae, the dispersed fringe visibilities and differential phases were obtained in spectral regions containing the H$\alpha$ and HeI 6678 lines and the H$\beta$ and HeI 4921 lines. 
Whereas the source is unresolved in the continuum, the source of the emission lines is resolved and the photocenter of the bulk of the H$\alpha$ emission exhibits offsets correlated with the orbital phase.  
For $\upsilon$~Sgr, both the continuum and H$\alpha$ sources are resolved, but no clear binary signal is detected. The differential phase shift across the line reveals that the bulk of the H$\alpha$ emission is clearly offset from the primary.
\end{abstract}


\section{Presentation of the CHARA array and the VEGA instrument}   
High angular resolution observations by optical long baseline interferometry, combined with classical  techniques like photometry and spectroscopy provide deep insight on the morphology of the massive interacting binaries and new constraints for their modeling.    
Here we present the first spectro-interferometric observations of the interacting massive binaries $\beta$ Lyr and $\upsilon$ Sgr carried out with the instrument VEGA on the CHARA interferometric array.     
The CHARA\footnote{http://www.chara.gsu.edu/CHARA/} interferometric array was constructed by the Center for High Angular Resolution (Georgia State University, Atlanta)(Brummelaar et al. 2005). Located at the Mount Wilson observatory (California), it is formed by six 1m telescopes placed by pairs on an Y shaped baseline. 
Base lengths of 34 m to 330 m allow to reach an angular resolution $\lambda / B$ up to $1.0~mas$ at near infrared wavelengths or $0.3~mas$ in the visible. 
The Visible spEctroGraph and polArimeter (VEGA) of the C\^{o}te d'Azur Observatory is a beam combiner operating in the visible range [$0.45-0.85 \mu m$] and offers three spectral resolutions ($R = 30000, 5000, 1700$). 
It is designed to record simultaneously the spectrally dispersed interference fringes with two photon counting detectors looking at two different spectral bands.
A polarimetric device can be inserted to record the polarized fringes. 
VEGA currently recombines 2 beams but 3 and 4 beams recombination is foreseen.
The data reduction process allows to measure the squared visibility modulus $V^2$ or the complex differential visibility $Ve^{i\varphi}$ as a function of the wavelength.
Spectro-interferometric observations with the VEGA instrument are dedicated to stellar physic programs for the determination of fundamental stellar parameters or the study of stellar activity and circumstellar environment.
A detailed description of the VEGA instrument is given by Mourard et al. (2008).   
\section{$\beta$ Lyrae}
$\beta$ Lyr (HD 174638) is a 12.9 d spectroscopic and eclipsing binary with evidence of circumstellar matter within and around the system.
First interferometric observations have been carried out in 1994 with the GI2T interferometer during the multi-technique campaign devoted to the study of this system (Harmanec et al. 1996).  
The combined analysis of the spectro-interferometric, spectroscopic and photometric observations lead to the conclusion that the source of the $H\alpha$ and $HeI$ emission lines can be associated with materials outside the orbital plane (jet like structure) of the binary.
In 2005, the first reconstructed images of the $H\alpha$ emission at low spectral resolution were obtained by differential phase referencing with the NPOI interferometer (Schmidtt et al. 2009).
The first resolved images of the core of the system, the eclipsing binary, were obtained in 2007 from near infrared observations (H band) with the MIRC beam combiner, thanks to the use of the longest baselines of the CHARA array. The elements of the relative orbit of $\beta$ Lyr have been derived from these interferometric observations (Zhao et al. 2008).                     

The observations of $\beta$ Lyr was dedicated to study of the morphology and the kinematic of the $H\alpha$ region. They were carried out during four nights in 2008 (orbital phases 0.5, 0.8, 0.9 \& 0.2) with the medium spectral resolution. 
For the observing dates the expected behavior of $\beta$\,Lyr can be predicted using the ephemeris from Ak et al. (2007) and the elements of the relative orbit (Zhao et al. 2008). 
The observations of $\beta$ Lyr were bracketed by these of the calibrator $\gamma$ Lyr.
With the 34 m S1S2 baseline, the angular resolution was 4.0 or 3.0 mas around $H\alpha$ and $H\beta$, respectively.
The direction of the baseline projected onto the sky were nearly perpendicular to the orbital plane.   
The data analysis have been done using the measurements of squared visibility modulus in continuum spectral bands and differential visibility measurements around the spectral lines $H\alpha$ and $HeI 6678$ ($R = 1650$) or $H\beta$ and $HeI 4921$ ($R = 1225$).

In the spectral continuum, the binary $\beta$ Lyr is unresolved as expected from the short baseline used and the orientation of the orbit.
In the $H\alpha$ line, at any orbital phase, the dip of the visibility modulus shows that the source is significantly resolved. The differential phase exhibits a signal in the line indicating a shift of the source photocenter correlated with the orbital phase. In $HeI 6678$ line, the source is marginally resolved at any orbital phases but no clear photocenter shift is detected.
For the $H\beta$ line, the source is resolved at some orbital phases and there is some suspicions of a photocenter shift. For the $HeI 4921$ line, there is no signal above the noise.

Work is in progress to determine the true extension of the $H\alpha$ emitting source and its location in the system.
These results will be compared with those of the observations of $\beta$~Lyr with NPOI (Schmitt et al. 2009), taking into account their low spectral resolution (R~$\sim$~40).   
   
\section{$\upsilon$ Sagittarii}
$\upsilon$~Sgr (HD 181615) is the brightest of the extremely hydrogen-deficient binaries stars (HdB stars), known as a 137.9~d spectroscopic binary. 
A new orbital solution and a better understanding of the the characteristics of this system were obtained from spectroscopic monitoring (Koubsk\'y et al. 2006).  
The presence of circumstellar matter in the $\upsilon$~Sgr system is evidenced by the complex H$\alpha$ absorption/emission profiles and by a strong infrared excess.
The first mid-IR interferometric observations of $\upsilon$~Sgr were carried out using the MIDI/VLTI instrument (Netolick\'y et al. 2009).
The dusty circumbinary disk is resolved in the N band with a typical size of 20 x 14~mas, an inclination $i \simeq 50 ^\circ$ and a position angle $PA \simeq 80 ^\circ$. Using these values to constraint the orbital parameters clearly confirm that $\upsilon$~Sgr is a massive binary with a total mass $> 15~M_\odot$.    

The observations of $\upsilon$~Sgr were carried out during three nights in 2008 (orbital phase 0.06-0.09) with the medium spectral resolution in spectral band centered on 660 nm. 
The appearance of $\upsilon$ Sgr (relative position of the components, location of the $H\alpha$ emission) at the observing dates can be predicted with some accuracy using the orbital elements (Koubsk\'y et al. 2006 and Netolick\'y et al. 2009). With a luminosity ratio $> 100$ the companion is expected to be almost unseen (Dudley \& Jeffery 1990).  
A 23 m S1S2 baseline ($\lambda/B \simeq 5.9~mas$) close to the N-S direction was mainly used and some data were also recorded with the E-W 100 m W1W2 baseline ($\lambda/B \simeq 1.4~mas$).
The observations of $\upsilon$~Sgr were bracketed by these of the calibrators $\gamma$~Sgr and $\upsilon$~Cap.
For the S1S2 baseline, the data analysis can been done using the differential spectral interferometric analysis (binned $R = 660$) around the $H\alpha$ line. Modulus visibility measurements were also done in the continuum ($\Delta\lambda=13~nm$ at $\lambda = 641, 656 ~\&~ 666~nm$) for the two baselines.

In the spectrum of $\upsilon$~Sgr, the $H\alpha$ line exhibits a `P Cygni' profile as expected at this orbital phase from Koubsk\'y et al. (2006).
For the two baselines, the source appears almost resolved in the continuum.
For the S1S2 baseline, a dip of the visibility associated with a shift of the differential phase of $\simeq 25^\circ$ in the $H\alpha$ line shows that the emitting source is well resolved and displaced from the continuum source.
Preliminary conclusions can be drawn from these first results.
About 10-20\% of the total flux may come from an over-resolved source and 90-80 \% from a compact source. The extended source could be related to the diffuse light from the inner rim of the dusty circumbinary disk. The compact source is probably dominated by the flux of the brigth component.
The source associated with the $H\alpha$ emission is resolved and its photocenter seems to be located between the two components as expected by Koubsky et al. (2006).
No sign of binarity is evidenced probably because of the difference of flux between components.  

Works are in progress to estimate the true $H\alpha$ extension and to convert the phase shift into photocenter position of the $H\alpha$ bulk emission. 

\section{Conclusion}
These first VEGA/CHARA observations of $\beta$~Lyr and $\upsilon$~Sgr show the potential of this instrument and the spectro-interferometry technique for the study of hot compact sources in general, and massive interacting binaries. To go further in the study of these complex systems, the next step will be to carry out observations at longer baselines to resolve the binary system and also multi-baselines observations to more stongly constrain the morphology of these sources.
\begin{figure}
\centering
\plotone{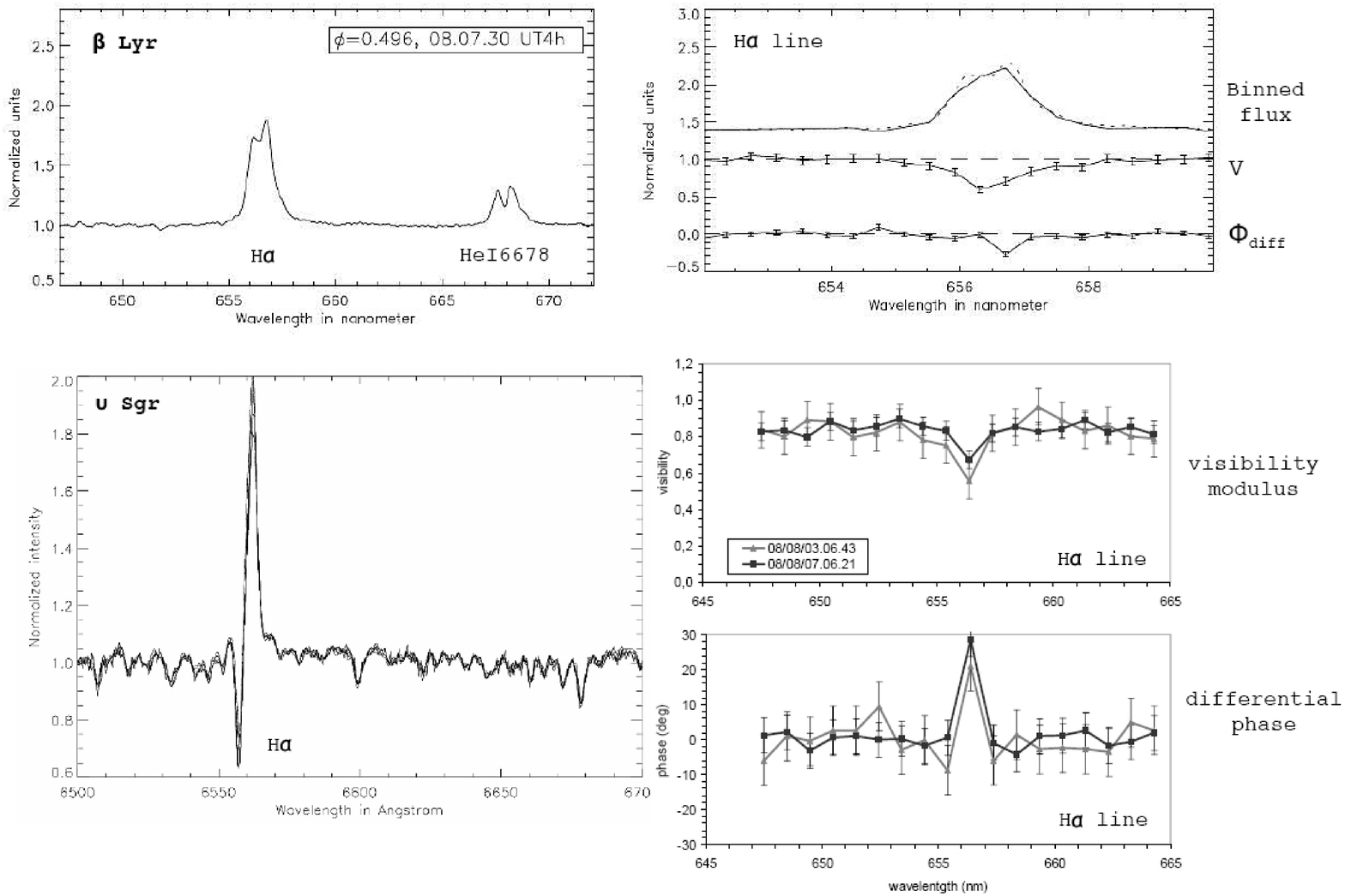}
\caption{Example of interferometric observations around H$\alpha$ line.
Top: $\beta$~Lyr. Bottom: $\upsilon$~Sgr
Left: spectrum recorded by VEGA. 
Right: result of the differential visibility analysis.
Note that the line intensity can be biased due to a local saturation effect of the photon counting camera.  
}
\label{fig_results}
\end{figure}
\acknowledgements 
VEGA/CHARA observations were possible thanks to the strong support of the CHARA team. Our participation at this conference have been supported by the CNRS international collaboration program ''France - R\'{e}publique Tch\`{e}que''(PICS 4343) and the French program for stellar physics (PNPS).  

%
%

%
\end{document}